\documentclass{eptcs}
\providecommand{\event}{PLACES 2016}

\usepackage[T1]{fontenc}
\usepackage[utf8x]{inputenc}
\usepackage[english]{babel}

\usepackage{latexsym}
\usepackage{hyperref}
\usepackage{courier}

\usepackage{amsmath,amsfonts,amssymb}
\usepackage{listings}
\usepackage{mathpartir}
\usepackage{mystyle}
\usepackage{xcolor}

\allowdisplaybreaks

\begin{document}
\title{Event-driven Adaptation in COP \thanks{Work partly supported by project PRA\_2016\_64 ``Through the fog'' funded by the University of Pisa.}}
\def\titlerunning{Event-driven Adaptation in COP}

\author{Pierpaolo Degano \qquad\qquad Gian-Luigi Ferrari\qquad\qquad Letterio Galletta \institute{Dipartimento di Informatica, Università di Pisa, Pisa, Italia.\\
\email{\{degano,giangi,galletta\}}@di.unipi.it}}

\def\authorrunning{Degano, Ferrari and Galletta}

\maketitle
\begin{abstract}
Context-Oriented Programming languages provide us with primitive constructs to adapt program behaviour depending on the evolution of their operational environment, namely the context.
In previous work we proposed \coda, a context-oriented language with two-components: a declarative constituent for programming the context and a functional one for computing.
This paper describes an extension of \coda\ to deal with adaptation to unpredictable context changes notified by asynchronous events.
\end{abstract}


%
         



\section{Introduction}
  \label{sec:intro}

\emph{Software is eating the world}: it is a fundamental component of our everyday objects, that all together form what is often called the \emph{Internet of Things}.
In this scenario, the physical resources (e.g.\ your coffeemaker) and the virtual ones (e.g.~your calendar) create a \emph{highly dynamic and open  virtual computing platform}, often referred to as the \emph{context}. 
The context ``virtualises''  the things, namely the \emph{smart} resources so as to make them appear less heterogeneous, unlimited and fully dedicated to their users.
Also, the context provides an abstract and uniform communication infrastructure to access smart things.
A distinguishing characteristic of smart resources is that in principle they are always connected to the Internet, possibly  through different access points.
Hence, they can interact with each other by collecting and exchanging information of various kinds and perform actions that modify the context.
For instance, your smart alarm clock can switch on your coffeemaker to prepare you a cup of coffee. 
In addition, they can choose on their own \emph{where, when} and to \emph{whom} they are visible and to \emph{which} portions of the context.

A key challenge is designing software systems that run without compromising their intended behaviour or 
their non-functional requirements, e.g.~quality of service, when injected in such highly dynamic and open contexts.
Programming these systems thus requires new programming language features and effective mechanisms 
to deal with \emph{context-awareness}, i.e.\ \emph{sensing}, \emph{reacting} and properly \emph{adapting} to changes of the actual context.
A comprehensive discussion on the software engineering challenges of the \emph{open world assumption} and of \emph{adaptation} is in~\cite{Baresi06}.

Using traditional programming languages, context-awareness is usually implemented by modelling the context through a special data structure, which can answer to a fixed number of queries, whose results can be tested through suitable $\myif$ statements.
This approach however charges the programmer with the responsibility of implementing this data structure and the corresponding operations. 
In addition, she is in charge of achieving a good modularisation, as well as a good separation of cross-cutting concerns, and of the interactions between the code using the context and legacy code.
More flexible proposals are specific design patterns~\cite{Ramirez:2010} and Aspect-Oriented Programming~\cite{KiczalesHHKPG01} that encapsulate context-dependent behaviour into separate modules, but still leave most of the burden of handling the context and its changes (and their correctness) to the programmers.
Recently, Context Oriented Programming (COP)~\cite{CostanzaH05} was proposed as a viable paradigm to develop context-aware software.
It advocates languages with peculiar constructs that express context-dependent behaviour in a modular manner.
In this way context-awareness is built-in, and the provided linguistic abstractions impose a good practice to programmers.
This makes a positive impact on the correctness and the modularity of code, mainly because low level details about context management are masked by the compiler.

We proposed in~\cite{ieee16} a programming language, called \coda, within the COP paradigm. 
\coda\ has two components: a logic constituent for programming the context and a functional one for computing.
The logical component provides high level primitives for describing and interacting with complex working environments.
Every resource in the context is described by a set of logical predicates representing its state.  
Thus a programmer can retrieve information from it by asking a query or she can use it by either asserting or retracting logical facts.
The functional component offers support for programming a variety of adaptation patterns. 
Its higher-order facilities are essential to exchange the bundle of functionalities required to manage adaptivity. 
\coda\ is equipped with a formal semantics which drove a prototypical implementation in F\#~\cite{foclasa15}.
Moreover, \coda\ offers a further support through a static analysis that guarantees programs to always be able to adapt in every context~\cite{ieee16}.

Our previous work focussed more on the foundational, linguistic and analysis aspects of context-aware programming, rather than on the way smart resources in the context interact and coordinate.
This paper takes a step towards a formal definition of coordination mechanisms.
Our context virtualises the smart resources and the communication infrastructure, as well as other software components running within it.
Consequently, the dynamic evolution of a context abstractly represents all the interactions of the entities it hosts.
It is enough therefore to consider a single application plugged in a context to capture all the interactions with the other entities therein.
Our formal coordination model is based on a notion of \emph{event}, thrown to notify when a relevant change occurs in the context.
In the wake-up example above, the application is the coffee machine, that will receive an event from the context, actually from the alarm clock, a few minutes before it rings. 

Technically, this work enriches the semantics of~\cite{ieee16} with mechanisms for throwing, receiving and handling events.
Since the context changes independently of the application, events are inherently asynchronous.
In our model applications are not strictly event-driven, and execute on their own, provided the resources they need are available.
However, the context evolution may make one of these resources unavailable, and upon receiving the related event  the application has to appropriately react.
Event handling thus deeply affects the run-time behaviour, the formal definition and the implementation of the adaptation constructs.
For example, if your application is sending a message through an access point, upon receiving an event of disconnection, it has to adapt its behaviour by connecting to another access point, if any, and resend the undelivered message.

\paragraph{Structure of the paper}
Section~\ref{sec:example} briefly surveys the execution model of \coda\ and introduces through a running example the main features of our proposal.
The operational semantics of our extension is 
\mbox{%
in Section~\ref{sec:lang}, and then we conclude and shortly discuss related and future work in the last section.
}

\section{An example}
  \label{sec:example}

In this section we illustrate and discuss our proposal,  by intuitively presenting the main features of \coda. 
In the following example, we consider a multimedia guide for museums as an instance of the Internet of Things scenario.
We assume the museum has a wireless infrastructure that provides communication facilities, i.e.\ an Intranet.
First a user registers at a desk, specifies her profile and gets credentials, then she connects to the museum Intranet to download the guide  for her smartphone. 

\paragraph{The museum context}
\lstset{language=Prolog,mathescape,basicstyle=\ttfamily\small,showstringspaces=false}
As said, the notion of \emph{context}, i.e.\ the environment where applications and smart objects run, is fundamental for adaptive software.
Abstractly, it could be considered as a heterogeneous collection of data coming from different sources and having different representations. 
In~\cite{ieee16} we adopted a declarative approach to deal with the context:
it is a knowledge base implemented as a Datalog program, following a well-studied approach~\cite{Tanca2010,Loke:2004}.
Hence, programmers can concentrate on \emph{what} information the context has to include, leaving to the runtime support \emph{how} this information is actually collected and managed.
In practice, an \coda\ context is a set of facts that predicate over a possibly rich data domain, and a set of logical rules to deduce further implicit properties of the context itself.
Adaptive programs can thus query the context and retrieve relevant information, by simply checking a Datalog goal.

Below, we consider a small portion of our guide in order to show how to describe the relevant contextual information declaratively.
The following snippet of code declares the pricing policy of the museum, that together with the user profile supports the emission of the appropriate ticket.
In our example, the first two clauses grant a reduced ticket to under 10 or over 65 users from a European country; the third clause does the same to students; additionally if they study art, entrance is free:
\begin{lstlisting}
ticket(reduced) $\leftarrow$ user_age(x), x < 10, user_country(y), Europe(y). 
ticket(reduced) $\leftarrow$ user_age(x), x > 65, user_country(y), Europe(y). 
ticket(reduced) $\leftarrow$ user_work(student). 
ticket(free) $\leftarrow$ user_work(student),  user_study(art). 
\end{lstlisting}                
The predicates \lstinline+user_age+, \lstinline+user_country+, \lstinline+user_work+ and \lstinline+user_study+ retrieve data from the user profile and we check if they satisfy the given constraints, e.g.\ \lstinline+x < 10+ or \lstinline+Europe(y)+.

\paragraph{Adaptation constructs}

\lstset{language=Caml,morekeywords={dlet, when,tell,retract},keywordstyle=\bfseries}

The functional part of the language provides two main mechanisms for programming adaptation.
The first is \emph{context-dependent binding} through which a programmer declares variables whose values depend on the context:
\lstinline+dlet x = e1 when goal in e2+ means that
the variable \lstinline+x+ (called \emph{parameter} hereafter) may denote different objects, with different behaviour depending on the different properties of the current context, expressed by \lstinline+goal+.

The second mechanism is based on \emph{behavioural variations}, chunks of code that can be activated depending on information picked up from the context, so as to dynamically adapt the running application.
It is a list of pairs \lstinline+goal.expression+ within curly parenthesis, similar to pattern-matching, that alters the control flow of applications according to which of its goals holds in the context.
Behavioural variations have parameters and are (high-order) values so facilitating programming dynamic adaptation patterns. 

Now, we show how behavioural variations express context dependency in our museum guide. 
Users can buy tickets through their smartphone, either via the museum web page or via a text message.
The preferred way is stored in the user's profile, together with additional information (e.g.\ her nationality).
The payment is implemented by the behavioural variation $\code{buy}$  with parameter $\code{user\_id}$ in the following snippet, where we use a sugared syntax of \coda:
\begin{lstlisting}
fun buyTicket (ticket_kind) = 
    let buy = (usr_id){
      $\leftarrow$ payByWeb.
        let c = getPage () in 
          sendData c  ticket_kind  usr_id
      $\leftarrow$ payByText.
        let c = getNumber () in 
	      sendText c  ticket_kind  usr_id
	} in
	#buy (get_usr_id ())
\end{lstlisting}
The function  \lstinline+buyTicket+ takes in input the kind of ticket (previously inferred querying the user profile),
and then applies the relevant case of the behavioural variation \lstinline+buy+ to the \lstinline+usr_id+ of the buyer 
(syntactically, we prefix this kind of application with \# to distinguish it from the standard functional one).
Assume the user pays via a text (the goal \lstinline+$\leftarrow$ payByText+ is satisfied), so the second case of the behavioural variation is selected,  but just before sending the text (\lstinline+sendText+ invocation), the signal of the mobile network is lost. 
This change of the context raises the event \lstinline+signalLost+ which is notified to the smartphone.
The running application must then adapt his behaviour to the asynchronous notification.
In our example, the guide reacts to the event
by re-executing the behavioural variation above  selecting another alternative to complete the payment, provided that the smartphone can connect to the WiFi.

Of course, the interaction with the Datalog context is not limited to queries, but one can change the knowledge base through \lstinline+tell+/\lstinline+retract+ operations that add/remove facts.
Modifying the context via these operations could raise specific events for other applications in the context: an event could be abstractly thought as an asynchronous sequence of \lstinline+tell+ and \lstinline+retract+.
In our example, the network failure is caused by another entity in the context through the operation \lstinline+retract(phone_on)+.

Note that event-driven adaptation cannot be rendered by using conditional statements alone, 
since the change (notified by an event) may falsify the guard of the already selected behavioural variation, e.g.\ because a resource becomes unavailable.
Consequently, mechanisms are in order to automatically re-adapt the application to the new context, if possible at all.

We now intuitively illustrate how behavioural variations work and how events are handled influencing adaptation.
As we will fomalise later on, the semantics of \coda\ is given through a transition system the configurations of which have the form $\langle q, \rho, C, p\rangle$ where $q$ is an event queue, $\rho$ is an environment for binding parameters, $C$ is (a simplified form of) the context and $p$ the running program; we also assume to have a list of event handlers $h$.
Consider the snippet above and suppose that its execution reaches the point in which the behavioural variation \lstinline+buy+ is applied, represented by the following configuration:
\[
\langle \epsilon, \rho,\, C,  \mathtt{\# buy (id) } \rangle
\]
where $\epsilon$ stands for the empty queue of events, the actual value of $\rho$ is immaterial, and the context $C$ contains the fact \lstinline+phone_on+ and \lstinline+id+ is the value returned by \lstinline+get_usr_id+.
When the behavioural variation is applied, the second case of \lstinline+buy+ is selected as said above.
A portion of the current configuration is stored because it can turn out to be useful in restoring the execution after the occurence of an event.
In particular, we store the successful goal of \lstinline+buy+, the environment $\rho$ and the behavioural variation application itself. 
After few execution steps we obtain the following configuration:
\[
\langle \epsilon, \rho,\, C,  \mathtt{sendText\ c\  ticket\_kind\ usr\_id} \rangle
\]
As described above the event \lstinline+signalLost+ occurs and the configuration evolves to 
\[
\langle \mathtt{signalLost}, \rho,\, C,  \mathtt{sendText\ c\  ticket\_kind\ usr\_id} \rangle
\]
The context becames  
$C' = C \setminus \{\mathtt{phone\_on}\}$ and  the event queue is not empty, so we run the handler $e = h(\mathtt{signalLost})$ which connects the smartphone to the WiFi.
The new configuration stores in the forth position that the program computation is suspended for the completion of the handler $e$: 
\[
\langle \epsilon, \rho,\, C',\,  [e]\,\mathtt{sendText\ c\  ticket\_kind\ usr\_id} \rangle
\]
When the handler terminates we retrieve the information stored when entering the behavioural variation, that is the triple
$(\leftarrow \mathtt{payByText},\,\rho,\,\mathtt{\# buy (id)})$.
Since the goal does not hold in $C'$ we apply again the whole behavioural variation reaching the configuration:  
\[
\langle \epsilon, \rho,\, C',\,  \mathtt{\# buy (id)} \rangle
\]
and the computation goes on by selecting the first alternative. 
Instead, if the goal had still held, the computation would have continued by invoking \lstinline+sendText+.

\section{The language}
  \label{sec:lang}

In this section we introduce our formal model by extending the semantics of \coda\ presented in~\cite{ieee16}.
As discussed above, the context ``virtualises'' \emph{smart} objects by providing them with a common interface and a communication infrastructure through which they can interact.
Since the interaction is performed by changing the context, in our semantics  we only consider the application in hand, its running context and the raised events.
The execution of an application and event notifications are asynchronous, and thus the application does not see the context changes caused by an event until it has been fully handled.
An application has event handlers for reacting to specific events. 
The execution of an event handler is atomic, i.e.\ the application cannot start handling a new event before  the running handler has completed. 

\paragraph{Syntax}
The syntax of the Datalog component is standard so here we assume it given~\cite{CeriDatalog}. The syntax of the functional part is the following where we denote with $F$ and $G$ Datalog facts and goals, respectively:
\begin{align*}
\rho \in PEnv & \quad  x,f \in Var \quad \dynvar{x} \in  DynVar \,(Var \cap DynVar = \emptyset) \quad  \alpha \in Event \\
Va ::= & G.e \mid G.e,Va \qquad \qquad \qquad \qquad
v  ::=  c \mid \lambda_{f} x. e \mid (x)\{Va\} \mid F \\
e ::= & v \mid x \mid \dynvar{x} \mid e_1\,e_2 \mid \mylet\,x\,=\,e_1\,\myin\,e_2 \mid 
\mydlet\, \dynvar{x}\, = e_1\, \mywhen\, G\, \myin\, e_2\mid \\
& \myif\,e_1\,\mythen\,e_2\,\myelse\,e_3 \mid  \mytell(e_1) \mid \myretract(e_1) \mid e_1 \cup e_2 \mid  \#(e_1, e_2) \mid aux\_e\\
h ::= & \alpha \Rightarrow e \mid h ; h \qquad \qquad \qquad p ::= \ee{e_1}{e_2} \mid e \qquad \qquad \qquad
aux\_e ::=  e_1^{\texpr{G}{\rho}{e_2}} \mid \barexpr{e}
\end{align*}
It is an extension of the one presented in~\cite{ieee16}: values are constants, functions
(here written in the standard $\lambda$-notation and in the example declared via the keyword \lstinline+fun+), behavioural variations and facts; expressions include standard ML constructs ($\mylet$, $\myif$, etc.), context-dependent binding ($\mydlet$), context updates and behavioural variation application and concatenation.
We also have \emph{parameters}, i.e.\ variables $\dynvar{x} \in  DynVar$ the value of which can only be known when the running context is fully set up.
The novelties of this work are \emph{event handlers} $h$ which manage an event $\alpha$ by running $e$, and two kinds of extended expressions, $p$ and $aux\_e$, discussed below and used in the semantic definitions.

\paragraph{Semantics}
For the Datalog evaluation we adopt the top-down standard semantics for stratified programs~\cite{CeriDatalog}. 
Given a context $C \in Context$ and a goal $G$, we write $C \vDash G\, with\, \theta$ to mean that there exists a substitution $\theta$, replacing constants for variables, such that the goal $G$ is satisfied in the context $C$.


The semantics of \coda\ is defined by two transition systems working in a master-slave manner.
They are inductively defined for expressions with no free variables, but possibly with free parameters, that take a value in an environment $\rho \in PEnv$, updated  through the standard operator $\rho[\dynvar{x} \mapsto b]$. 
Actually, we assume that the arrival of an event $\alpha$ transforms a context $C$ in $C'$, written $C \xrightarrow{\alpha} C'$; here we are not  interested in how this happens, so this additional transition system is left abstract.
%

The master transition system models receiving (i.e.\ queuing) and handling events.
Its transitions have the form $\langle q, \rho, \mcxt, p\rangle \doublearrow \langle q', \rho', \mcxt', p'\rangle$ where $h$ is a list of event handlers, $q$ is an event queue, $\rho$ is an environment, $C$ is the context and $p$ the running program. 
By an abuse of notation, we use $h$ as a function from events to expressions: $h(\alpha) = ()$ when there is no handler for $\alpha$, otherwise $h$ returns the expression to run. 
We denote with $\epsilon$, $\alpha \cdot q$ and $q \cdot \alpha$, resp., the empty queue, a queue with front event $\alpha$, the queuing of $\alpha$, resp. 
The annotation $\bullet \in \{\evs, -\}$ on the context $C$ is a flag used to signal an event notification to the slave transition system; 
if $\bullet = \evs$, the event may affect the application execution, otherwise it is harmless.
The rules of the master transition system are the following:
{\small
\begin{mathpar}
\inferrule[Enew]{ }{\langle q, \rho, \mcxt, p\rangle \doublearrow \langle q \cdot \alpha, \rho, \mcxt, p\rangle}

\inferrule[Eman]{ }{\langle \alpha \cdot q, \rho, \mcxt, e\rangle \doublearrow \langle q, \rho, \emcxt', \ee{e'}{e}\rangle} \quad h(\alpha) = e' \quad C \xrightarrow{\alpha} C'


\inferrule[Ehdr1]{ \langle \rho, \mcxt, e_1\rangle \rightarrow \langle \rho', \mcxt', e'_1\rangle }{\langle q, \rho, \mcxt, \ee{e_1}{e_2}\rangle \doublearrow \langle q, \rho', \mcxt', \ee{e'_1}{e_2}\rangle}

\inferrule[Ehdr2]{ }{\langle q, \rho, \mcxt, \ee{()}{e}\rangle \doublearrow \langle q, \rho, \mcxt, e\rangle}

\inferrule[Eexpr]{ \langle \rho, \mcxt, e\rangle \rightarrow \langle \rho', \mcxt', e'\rangle }{\langle \epsilon, \rho, \mcxt, e\rangle \doublearrow \langle \epsilon, \rho', \mcxt', e'\rangle}
\end{mathpar}
}
The rule \rulename{Enew} queues a new event $\alpha$ in $q$, upon arrival; it resembles an early input rule, and it is always enabled.
The rule \rulename{Eman} dequeues an event $\alpha$ from $q$, updates the context ($C \xrightarrow{\alpha} C'$) and launches the corresponding event handler $h(\alpha)$, recording in the extended expression $[e']e$ both the handler and the suspended application. 
Since $[e']e$ is not an expression $e$, we guarantee the atomicity of event handlers.
We also assume that event handlers do not perform long run tasks, and that always terminate.
The rules \rulename{Ehdr1} and \rulename{Ehdr2} run an event handler (within an extended expression), while the rule \rulename{Eexpr} runs an expression. 
Note that this rule can only be used when the event queue is empty, ensuring that the application runs in a context which is ``stable''.

We now briefly present the rules of the slave transition system, starting from the ones for the usual ML constructs.
They are quite standard (just remove the context $\mcxt$), so we do not comment on them:
%

\begin{mathpar}
\small
\inferrule[If1]{\langle \rho, \mcxt,\,e_1\rangle \rightarrow \langle \rho', \mcxt',\,e'_1\rangle}{\langle \rho, \mcxt,\,\myif\,e_1\,\mythen\,e_2\,\myelse\,e_3\rangle \rightarrow \langle \rho', \mcxt',\,\myif\,e'_1\,\mythen\,e_2\,\myelse\,e_3\rangle}

\inferrule[If2]{ }{\langle \rho, \mcxt,\,\myif\,true\,\mythen\,e_2\,\myelse\,e_3\rangle \rightarrow \langle \rho, \mcxt,\,e_2\rangle}

\inferrule[If3]{ }{\langle \rho, \mcxt,\,\myif\,false\,\mythen\,e_2\,\myelse\,e_3\rangle \rightarrow \langle \rho, \mcxt,\,e_3\rangle}

\inferrule[Let1]{\langle \rho, \mcxt, e_1 \rangle \rightarrow \langle \rho', \mcxt', e'_1 \rangle}{\langle \rho, \mcxt, \mylet\, x\, =\, e_1\, \myin\, e_2 \rangle \rightarrow \langle \rho', \mcxt', \mylet\, x\, =\, e'_1\, \myin\, e_2 \rangle}

\inferrule[Let2]{ }{\langle \rho, \mcxt, \mylet\, x\, =\, v\, \myin\, e_2 \rangle \rightarrow \langle \rho, \mcxt, e_2\{v/x\} \rangle}

\inferrule[App1]{\langle \rho, \mcxt,\,e_1 \rangle \rightarrow \langle \rho', \mcxt',\,e'_1\rangle}{\langle \rho, \mcxt,\,e_1\,e_2 \rangle \rightarrow \langle \rho', \mcxt',\,e'_1\,e_2\rangle}

\inferrule[App2]{\langle \rho, \mcxt,\,e_2 \rangle \rightarrow \langle \rho',\, \mcxt',\,e'_2\rangle}{\langle \rho, \mcxt,\,(\lambda_f x.e)\,e_2\rangle \rightarrow \langle \rho', \mcxt',\,(\lambda_f x.e)\,e'_2\rangle}

\inferrule[App3]{ }{\langle \rho, \mcxt,\,(\lambda_f x.e)\,v\rangle \rightarrow \langle \rho,\,\mcxt,\,e\{v/x, (\lambda_f x.e)/f\}\rangle}
\end{mathpar}


\medskip
The rules that handle the context are taken from~\cite{ieee16} and are as follows:
%
{\small
\begin{mathpar}

\inferrule[Tell1]{\langle \rho, \mcxt,\,e\rangle \rightarrow \langle \rho', \mcxt',\,e'\rangle}{\langle \rho, \mcxt,\, \mytell(e)\rangle \rightarrow \langle \rho', \mcxt',\,\mytell (e')\rangle}

\inferrule[Tell2]{ }{\langle \rho, \mcxt,\, \mytell(F)\rangle \rightarrow \langle \rho, \mcxt \cup \{F\},\,()\rangle}

\inferrule[Retract1]{\langle \rho, \mcxt,\,e\rangle \rightarrow \langle \rho', \mcxt',\,e'\rangle}{\langle \rho, \mcxt,\, \myretract(e)\rangle \rightarrow \langle \rho', \mcxt',\,\myretract (e')\rangle}

\inferrule[Retract2]{ }{\langle \rho, \mcxt,\, \myretract(F)\rangle \rightarrow \langle \rho, \mcxt \setminus \{F\},\,()\rangle}
\end{mathpar}
}
{\small
\begin{mathpar}
\inferrule[Dlet1]{ }{\langle \rho, \mcxt, \mydlet\, \dynvar{x} \, = \, e_1 \,\mywhen\, G_1\,\myin\, e_2\rangle \rightarrow \langle \rho[\dynvar{x} \mapsto G_1.e_1,\rho(\dynvar{x})], \mcxt', \barexpr{e_2}\rangle}

\inferrule[Dlet2]{\langle \rho, \mcxt, e\rangle \rightarrow \langle \rho', \mcxt', e' \rangle}{\langle \rho, \mcxt, \barexpr{e}\rangle \rightarrow \langle \rho', \mcxt', \barexpr{e}' \rangle }

\inferrule[Dlet3]{ }{\langle \rho[\dynvar{x} \mapsto Va], \mcxt, \barexpr{v}\rangle \rightarrow \langle \rho, \mcxt, v \rangle }

\inferrule[Append1]{\langle \rho, \mcxt,\,e_1\rangle \rightarrow \langle \rho', \mcxt',\,e'_1\rangle}{\langle \rho, \mcxt,\,e_1 \cup e_2\rangle \rightarrow \langle \rho', \mcxt',\,e'_1 \cup e_2\rangle}

\inferrule[Append2]{\langle \rho, \mcxt,\,e_2\rangle \rightarrow \langle \mcxt',\,e'_2\rangle}{\langle \rho, \mcxt,\,bv \cup e_2 \rangle \rightarrow \langle \rho', \mcxt',\,bv \cup e'_2 \rangle}

\inferrule[Append3]{ }{\langle \rho, \mcxt,\,(x)\{Va_1\} \cup (y)\{Va_2\}\rangle \rightarrow \langle \rho, \mcxt,\,(z)\{Va_1\{z/x\},\,Va_2\{z/y\}\}} \quad \text{if } z \text{ fresh }

\end{mathpar}
}
%
The application can change the context by asserting or removing facts, as shown by the rule \rulename{Tell2}, after the expression has been reduced first to a fact, by the inductive rule  \rulename{Tell1}.
%
The rules \rulename{Dlet} deal with context-dependent binding. 
The rule \rulename{Dlet1} extends the environment $\rho$ by appending $G_1.e_1$ in front of the existing binding for $\dynvar{x}$ ($e_1$ is \emph{not} evaluated until the running context is fully known). 
In addition, it overlines $e_2$ , so as to record that it can be evaluated in a context where $G_1$ holds, under the updated environment.
The other two rules are standard: the $\mydlet$ inductively reduces to the plain value eventually computed by 
$\overline{e_2}$.

Finally, the rules for $e_1 \cup e_2$ sequentially evaluate $e_1$ and $e_2$ until they reduce to behavioural variations (rules \rulename{(Append1, 2)}). 
Then, they are concatenated together by renaming bound variables to avoid name captures (rule \rulename{(Append3)}).

The rules for adaptation are as follows:
{\small
\begin{mathpar}

\inferrule[VaApp1]{\langle \rho, \mcxt,\,e_1\rangle \rightarrow \langle \rho', \mcxt',\,e'_1\rangle}{\langle \rho, \mcxt,\,\#(e_1,\,e_2)\rangle \rightarrow \langle \rho', \mcxt',\#(\,e'_1,\,e_2)\rangle}

\inferrule[VaApp2]{\langle \rho, \mcxt,\,e_2\rangle \rightarrow \langle \rho', \mcxt',\,e'_2\rangle}{\langle \rho, \mcxt,\,\#(bv,\,e_2)\rangle \rightarrow \langle \rho', \mcxt',\#(bv,\,e'_2)\rangle}

\inferrule[VaApp3]{ dsp(C, Va) = (e', G)}{\langle\rho,\mcxt, \#((x)\{Va\},v)\rangle \rightarrow \langle \rho, \mcxt, e'\{v/x\}^{\texpr{G}{\rho}{\#((x)\{Va\},v)}}\rangle } 

\inferrule[Dynvar]{ \rho(\dynvar{x}) = Va \\ dsp(C, Va) = (e', G)}{\langle \rho, \mcxt, \dynvar{x}\rangle \rightarrow \langle \rho, \mcxt, e'^{\texpr{G}{\rho}{\dynvar{x}}}\rangle}
\end{mathpar}
}%

The rules \rulename{VaApp} handle behavioural variations, that are alike abstractions and thus are applied.
The first two rules are standard and the third makes use of the so-called \emph{dispatching} mechanism to select the expression to which the behavioural variation reduces.
This inspects $\mathit{Va}$, i.e.\ a list of pairs ($G, e$), to find the first goal $G$ satisfied by the current context $C_\bullet$, under a substitution $\theta$ that binds the variables of $G$. 
An adaptation failure occurs if no such goal exists.
Otherwise, the result is the pair ($e' = e\theta, G$) and the behavioural variation reduces to the extended expression 
$e'\{v/x\}^{\texpr{G}{\rho}{\#((x)\{Va\},v)}}$.
The triple records all the information needed to restart the execution of the variation $Va$ if the occurrence of an event makes the goal $G$ no longer satisfiable.
Since goals are used to express those properties that select the appropriate case of behavioural variations, if such a situation arises the computation becomes unreliable, e.g.\ because some resource disappeared.
It is convenient then to restart the whole behavioural variation to recover from a possible failure, without undoing the work done so far.
A similar situation arises when a parameter $\dynvar{x}$ has to be evaluated.
From the variation $\mathit{Va}$ bound to $\dynvar{x}$ in $\rho$, through  \emph{dsp} the rule \rulename{Dynvar} selects an expression to which $\dynvar{x}$ reduces, if any, and stores the configuration from which to restart the computation if an event makes the context unstable.
The rules that handle recovery are:
{\small
\begin{mathpar}

\inferrule[Brk1]{ \langle \rho, \nmcxt, e \rangle \rightarrow  \langle \rho', \mcxt', e' \rangle}{ \langle \rho, \nmcxt, e^{\texpr{G}{\rho''}{e''}}\rangle \rightarrow \langle \rho', \mcxt', e'^{\texpr{G}{\rho''}{e''}}\rangle}

\inferrule[Brk2]{ }{ \langle \rho, \mcxt, v^{\texpr{G}{\rho''}{e''}}\rangle \rightarrow \langle \rho, \mcxt, v\rangle}

\inferrule[Brk3]{ C \vDash G \\ \langle \rho, \emcxt, e \rangle \rightarrow  \langle \rho', \mcxt', e' \rangle}{ \langle \rho, \emcxt, e^{\texpr{G}{\rho''}{e''}}\rangle \rightarrow \langle \rho', \nmcxt', e'^{\texpr{G}{\rho''}{e''}}\rangle}

\inferrule[Brk4]{ C \nvDash G }{\langle \rho, \emcxt, e^{\texpr{G}{\rho'}{e''}}\rangle \rightarrow \langle\rho, \nmcxt, e'' \rangle}

\end{mathpar}
}%
The actual recovery is specified by the rule \rulename{Brk4} that restores the situation \emph{ex quo ante}.
Note that the index of $C$ is $\evs$ showing that the event affects the context. 
Rule \rulename{Brk1} simply evaluates $e$, while \rulename{Brk3} says that the event does not affect the satisfiablity of the goal $G$ (note that the index $\evs$ becames $\_$).
When a value is reached, annotations are discarded (\rulename{Brk2}).

\paragraph{Discussion}
Our programming model can be ideally seen as composed by two threads, that operate in an interleaving fashion.
The first is the application one, and the other encapsulates, through the context, all the other entities within it. 
The interleaved execution gives raise to non-determinism, especially because the occurrence of events is asynchronous.
Our proposal guarantees that each event is handled atomically: before considering a new event, the execution of the handler of the previous one has to be terminated.
We assume to have no control on non-determinism, and thus an application may starve, when it spends all its time in handling  events incoming one right after the other.
A simple improvement would be assigning a priority to events, having in mind that an over-flood of events is infrequenty, especially of those with high priority.
Finally, note that here we are not specifying any compensation, as our goal is only to guarantee that the application runs in a context with all the needed features.
For taking into account compensations, it will be sufficient to enrich the auxiliary expression $aux\_e$ and modify and extend accordingly the group of rules \rulename{(Brk)}.

\section{Concluding remarks}
  \label{sec:concl}

We extended the COP language \coda\ to manage asynchronous changes of a context caused by the occurrence of events.
Some events may drive the execution of a behavioural variation into an unreliable state, e.g.\ when a device  accessed in the current run is switched off.
Our semantics supports recovery in these cases.
A non trivial effort is needed to extend the static analysis of~\cite{ieee16} that guarantees adaptations to be always successful.
We plan to carry on this check at run-time, when a behavioural variation is about to run.
If successful, such check will avoid at all running a recovery procedure.
Also, the \coda\ prototype will be extended to deal with asynchronous events.

\paragraph{Related work}
As far as we know a limited number of papers in COP literature has considered event-driven adaptation, among which we briefly mention those with features similar to ours.
EventCJ~\cite{Aotani:2011} is a Java-based language which combines mechanisms from COP with event based context changes. 
Differently from our proposal, it provides constructs to declare both the events thrown by an application and the transition rules specifying how to change the context when an event is received.
The formal semantics of ContextErlang in~\cite{SalvaneschiGP15} describes the 
behaviour of the constructs for adaptation within a distributed and concurrent framework.
The messages of ContextErlang are similar to the our asynchronous event, and so is their handling.
The language Flute~\cite{engineer:2012} is designed for programming reactive adaptive software. 
Flute constrains the execution of a procedure with certain contextual properties specified by a developer.
If any of these properties is no longer satisfied, the execution is stopped, until the property holds again.
Stopping the execution is like in our approach when goal of a behavioural variation has been falsified because of a context change.
However, here we proposed a different restart mechanism, based on the recovery specified by the rules \rulename{Brk}.

\bibliographystyle{eptcs}
\bibliography{biblio}

%
   
\end{document}